\documentclass[aps, amsfonts, nofootinbib,notitlepage, preprintnumbers]{revtex4-1}
\usepackage{hyperref}
\usepackage{ mathrsfs }
\hypersetup{
colorlinks=true,
linkcolor=red,
citecolor=blue,
urlcolor=blue}
\usepackage{epsf,epsfig}
\usepackage{amscd}
\usepackage{amsmath}
\usepackage{subfig}
\usepackage{slashed}
\usepackage{graphicx}
\usepackage[countmax]{subfloat}
\input xy
\xyoption{all}

\newcommand{\bea}{\begin{eqnarray}}
\newcommand{\eea}{\end{eqnarray}}

\begin{document}

\title{Anisotropic Propagator for the Goldstone Modes in Color-flavor Locked Phase in the Presence of a Magnetic Field} 
\author{Srimoyee Sen\footnote{{\tt srimoyee@umd.edu}}}
\affiliation{Maryland Center for Fundamental Physics,\\ 
Department of Physics,\\ 
University of Maryland, College Park, MD USA}

\preprint{UM-DOE/ER/...}

\begin{abstract}
We consider the phase diagram of QCD at very high baryon density and at zero temperature in the presence of a strong magnetic field. The state of matter at such high densities and low temperatures is believed to be a phase known as the color-flavor locked phase which breaks color and electromagnetic gauge invariance leaving a linear combination of them, denoted as $U(1)_{\tilde{em}}$ unbroken. Of the $9$ quarks (three flavors and three colors), five are neutral under this unbroken generator and four are oppositely charged(two with a charge of $+1$ and two with $-1$). In the presence of a magnetic field corresponding to $U(1)_{\tilde{em}}$ however, the properties of the condensate changes and a new phase known as the magnetic color flavor locked  (MCFL)phase is realized. This phase breaks an approximate $SU(2)_{C}\times SU(2)_{L}\times SU(2)_{R}\times U(1)_B\times U(1)_A$ symmetry of the Lagrangian to $SU(2)_{C+L+R}$ giving rise to $6$ Goldstone modes $5$ of which are pseudo Goldstone modes. These Goldstone modes are composed of excitations that correspond to both neutral quarks and charged quarks. Hence it is natural to expect that the propagators of these Goldstone modes get affected in the presence of a magnetic field and their speed becomes considerably anisotropic. Although this anisotropy is self-evident from symmetry arguments, it has not been quantified yet. We calculate this anisotropy in the speed of the Goldstone modes using an NJL model type of interaction between the quarks and comment on the impact of such anisotropic modes on transport properties of the MCFL phase.
	
\end{abstract}
\maketitle

\section{Introduction}
The low energy properties of any state of matter is primarily dictated by its lowest energy excitations. In case of a Bose condensate these excitations are usually the gapless Goldstone modes which arise due to the spontaneous breaking of continuous symmetries by the condensate. This paper explores the effect of a magnetic field on the propagation of the Goldstone modes that arise due to the spontaneous breaking of color-flavor symmetries in a condensed phase of matter known as the color-flavor locked (CFL) phase in the presence of a magnetic field. Before going into the details of what color-flavor locking is we can expect to predict some of the consequences of having a magnetic field using symmetry arguments alone. The presence of a magnetic field breaks rotational symmetry explicitly and the most natural consequence of it on the propagator of the Goldstone modes can be a difference in the speed of the modes in a direction perpendicular to the field compared to a direction parallel to the field. Although the effective theory of the Goldstone modes of the CFL phase in the presence of a magnetic field has been discussed in some detail in \cite{PhysRevD.76.045011}, the effect of a magnetic field on the speed has been commented on at a qualitative level. We attempt to quantify the effect of the magnetic field on the neutral Goldstone mode propagators of color-flavor locked phase in this paper.\\\\
\indent To put things in context, the phase diagram of QCD at finite baryon densities garners a lot of attention due to its implications for the physics of compact astrophysical objects, terrestrial nuclear experiments and heavy ion collisions. Although considerable progress has been made in exploring the phase diagram, some parts of it still remain inaccessible to rigorous analytical or numerical treatment. The region of the phase diagram that corresponds to moderate to very high baryon densities at low temperature is particularly relevant for the physics of compact stars, the reason being that these stars are cold and composed mostly of baryons. However, the regime with moderate baryon density and low temperature is difficult to analyse due to the following reasons. The strength of the QCD coupling constant at low energies being large renders perturbative calculations from first principles futile necessitating the use of numerical calculations on lattice. But lattice simulations at finite baryon density are not useful either as the lattice algorithms involving important sampling break down due to the sign problem\cite{Nicholson:2012xt,deForcrand:2008vr,Fodor:2001pe,Ejiri:2003dc,PhysRevD.71.114014}. 
However there is a region of the phase diagram, at very high baryon densities and low temperatures, that can be dealt with analytically. This can be explained as follows. At high density and low temperature quarks form Fermi spheres. The physics, in this regime of the phase diagram, is mostly dictated by the quarks at the Fermi surface. As the fermions at the Fermi surface are highly energetic, they are weakly coupled making rigorous perturbative calculations justified. This regime also happens to be relevant for the physics of the core of neutron stars as the density of baryons at the core is expected to reach asymptotically high values. The high density regime has been explored in great detail in the past two decades. In this regime matter was predicted to exist in a colorsuperconducting phase known as the `color-flavor locked' phase using both model calculations \cite{Alford:1997zt,PhysRevLett.81.53,PhysRevD.60.016004,Buballa:2001gj,PhysRevC.65.045204,PhysRevD.65.076012} and weak coupling calculations in QCD \cite{PhysRevD.59.094019,PhysRevD.61.056001,PhysRevD.60.114033,Hsu:1999mp,PhysRevD.61.051501}, \cite{,Shovkovy:1999mr,Schafer:2003jn,PhysRevD.69.074012} and \cite{PhysRevD.71.054003,PhysRevLett.93.232301,PhysRevD.73.094009}. The term `color superconductivity' corresponds to the Bose condensation of color-charged Cooper pairs of two quarks at the Fermi surface. As mentioned before at high densities and low temperatures quarks form Fermi spheres with a radius that depends on the baryon chemical potential. In such a scenario, if there is an attractive color-flavor and angular momentum channel available at the Fermi surface giving rise to a BCS instability, two quarks at the Fermi surface can form a Cooper pair lowering their energies. The Cooper pairs being Bosonic excitations, form a Bose-Einstein condensate which is not color neutral. The gluons acquire a Meissner mass in such a condensate and get screened. Many such color superconducting phases with different quark pairings have been proposed for a range of baryon densities. At asymptotically high densities the energetically favoured color superconducting phase was found to be the color-flavor locked phase. The condensate is not color neutral as expected and breaks both color and $U(1)_{em}$. However, there is a linear combination of the $8^{th}$ generator of the color $SU_c(3)$ and the $U(1)_{em}$ that remains unbroken by the condensate. This linear combination is referred to as the rotated electromagnetism and denoted by $U(1)_{\tilde{em}}$. A magnetic field corresponding to this rotated electromagnetism ($\tilde{B}$) does not get screened by the condensate. The scenario at this point diverges markedly from the $U(1)_{em}$ superconductivity in terrestrial metals which shields magnetic fields completely or accomodates magnetic fields only inside quantized vortices. This opens up exciting new possibilities as far as the properties of CFL phase in the presence of a magnetic field is concerned. But large magnetic fields in CFL phase are  interesting not for theoretical reasons alone. There is usually a very strong magnetic field associated with a compact star. For a neutron star the typical fields on the surface are such that $B\sim 10^{12}$ Gauss and for a magnetar the fields could be as large as $10^{16}$ Gauss\cite{Fushiki,RevModPhys.73.629,0004-637X-473-1-322}. For gravitationally bound stars an upper bound on the strength of the magnetic field can be found by comparing the energy of the gravitaional field and that of the magnetic field of the star. This bound is around $10^{18}$ Gauss. If we consider self bound quark stars however, this upper limit can be even higher\cite{PhysRevD.76.105030}. Also, higher magnetic fields are expected to exist inside $U(1)_{\tilde{em}}$ charged gluonic vortices \cite{Ferrer:2013txa,PhysRevD.76.114012} which arise due to chromomagnetic instability in dense matter\cite{PhysRevD.70.051501,PhysRevD.70.094030,Huang:2003xd} in the presence of a magnetic field. Typical magnetic field strength in the core of magnetars may well reach the order of the square of the CFL gap or higher and can be expected to alter the nature of the diquark pairing of the condensate giving rise to a different color superconducting phase known as the magnetic CFL or MCFL phase.    
\begin{center}
\begin{table}
\label{tab}
\begin{tabular}{ |c|c|c|c|c|c|c|c|c| } 
 \hline
 $s_b$ & $s_g$ & $s_r$ & $d_b$ & $d_g$ & $d_r$ & $u_b$ & $u_g$ & $u_r$ \\ \hline
 0 & 0 & -1 & 0 & 0 & -1 & 1 & 1 & 0  \\ \hline
 \end{tabular}
\caption{The $U(1)_{\tilde{em}}$ charge of the quarks}
\label{tab}
\end{table}
\end{center}
The symmetry properties of the magnetic CFL phase are slightly different from that of the CFL phase. To understand what these differences are let us first look at the symmetry breaking patterns of the CFL phase in the absence of a magnetic field. The standard model Lagrangian with a baryon chemical potential has approximate $SU(3)_c\times SU(3)_{L}\times SU(3)_R\times U(1)_B\times U(1)_A$ symmetry well below the weak scale. The condensation of CFL phase breaks it to $SU(3)_{C+L+R}$ giving Meissner mass to seven of the eight gluons. A linear combination of the $8^{th}$ gluonic generator and the generator of $U(1)_{em}$ also becomes massive where as the orthogonal linear combination of the two remains massless ($U(1)_{\tilde{em}}$). The CFL phase has a total of $10$ Goldstone modes including $\pi^{\pm}, \pi^0, K^0, \bar{K}^0, K^{+}, K^{-}, \eta, \eta'$ mesons in the massless quark limit. In addition there is a Goldstone mode due to the breaking of $U(1)_{B}$ which we denote by $\phi$. In order to predict the observable consequences, if any, of the existence of a CFL phase in the core of a neutron star it is of utmost importance to analyse the effective theory of these Goldstone modes. The low energy effective theories of these Goldstone modes were discussed in great detail in \cite{Beane:2000ms,Casalbuoni:1999wu,Hong:1999ei,PhysRevD.61.074012,Rho:1999xf,Manuel:2000wm} more than a decade ago. The effect of a $U(1)_{\tilde{em}}$ magnetic field $\tilde{B}$ on these Goldstone modes however remains to be explored. But in order to discuss the effects of $\tilde{B}$ on these Goldstone modes adequately we need to also take into account how the pairing of the condensate and consequently the symmetry properties get affected by this magnetic field. In the presence of $\tilde{B}$, the aforementioned $SU(3)_c\times SU(3)_{L}\times SU(3)_R\times U(1)_B\times U(1)_A$ symmetry of the standard model Lagrangian gets explicitly broken down to $SU(2)_c\times SU(2)_{L}\times SU(2)_R\times U(1)_B\times U(1)_A$ symmetry as four of the nine quarks are electrically charged under this $U(1)_{\tilde{em}}$. The charges of the quarks are shown in Table. \ref{tab}. When the magnetic field is small compared to the CFL gap the effect of the magnetic field on the structure of the gap is negligible and the form of the condensate is given by that of the CFL pairing. However, when the magnetic field becomes comparable to the CFL gap, the form of the quark-pairing changes and the condensate breaks $SU(2)_c\times SU(2)_{L}\times SU(2)_R\times U(1)_B\times U(1)_A$ down to $SU(2)_{R+L+C}$\cite{Ferrer:2006vw,PhysRevLett.95.152002,PhysRevD.76.105030}. The gap in the MCFL phase was calculated for both magnetic fields that are extremely large ($\mu^2= \tilde{e}\tilde{B}$, $\sim10^{19}$ Gauss for $\mu\sim 500$MeV) \cite{Ferrer:2006vw,PhysRevLett.95.152002,PhysRevD.76.105030} and for smaller magnetic fields ($10^{19}\text{Gauss}>\tilde{e}\tilde{B}>10^{18}$ Gauss)\cite{PhysRevD.76.105030}. It was also shown that if the magnetic field is lowered further the gap structure of the magnetic CFL phase becomes identical to that of CFL. The lower limit  Four of the Goldstone modes of the CFL phase are massive in the MCFL phase due to the explicit breaking of some of the color-flavor symmetry of the QCD Lagrangian. These are the modes that are charged under the $U(1)_{\tilde{em}}$ and are basically the charged pions and kaons. The MCFL phase retains the other $6$ Goldstone modes of the CFL phase listed above. The effective theory of these Goldstone modes was discussed to some extent in \cite{PhysRevD.76.045011}. These six modes are neutral under the rotated $U(1)_{\tilde{em}}$ and are made up of either two neutral quark excitations or two oppositely charged quark excitaions so as to keep the combined excitaion neutral. Small magnetic fields do not affect the Goldstone mode propagation appreciably as the modes are neutral under this $U(1)_{\tilde{em}}$. However a large magnetic field $\tilde{B}$ should be able to resolve the internal structure of the modes and be able to change their properties. One such effect is encapsulated in the speeds of the Goldstone modes and we can quantify this effect by looking at the underlying theory from which the effective theory of the Goldstone modes is derived. In the presence of a magnetic field $\tilde{B}$ the quarks are basically stuck in Landau levels which not only alters the gap structure of the MCFL phase from the CFL phase but also causes the speeds of the neutral Goldstone modes to be different in directions parallel to the magnetic field compared to that perpendicular to it. The six neutral Goldstone modes are of interest to us as being neutral they are massless in the presence of a magnetic field and easy to excite at temperatures much below the gap. This means that the low energy properties of the MCFL phase are dictated by these neutral Goldstone modes. 
\\\\
The paper is organized as follows. In section Sec. \ref{magmu} we analyse magnetic fields as high as $\tilde{e}\tilde{B}\sim 2\mu^2$ and in section Sec. \ref{magD} we look at more realistic smaller magnetic fields followed by a concluding section. Although $\tilde{e}\tilde{B}\sim 2\mu^2$ may be too high a magnetic field to be present at the core of a compact star, it is not ruled out and may well be within the range of realistic magnetic fields in quark stars. Also, a mechanism was conjectured \cite{Ferrer:2013txa,PhysRevD.76.114012}by which high magnetic fields can be found to exist inside gluonic vortices in CFL phase. The mechanism is as follows. Under the rotated $U(1)_{\tilde{em}}$, some of the gluons are charged. This causes the dispersion relation of one of the polarizations of these charged gluons to become tachyonic for high magnetic fields creating a chromomagnetic instability. As a consequence charged gluons condense in a vortex state with magnetic flux tubes inside, which correspond to stronger magnetic fields than the magnetic field present outside the vortices. This phenomenon is called antiscreening of the magnetic field and can give rise to magnetic fields in the core of magnetars that are stronger than expected otherwise. Despite the magnetic field being too large, it is worth analysing such fields as the computation of the propagator for the neutral Goldstone modes is simplified and enlightening for the purpose of demonstrating how a similar calculation will work for smaller magnetic fields. 
\section{Goldstone mode propagator at very high magnetic fields}\label{magmu}
Before we compute the propagators for the Goldstone modes, the energy scales of interest need to be specified. Here, we are interested in the neutral gapless excitations about the MCFL condensate with momentum much smaller campared to the gap which is much smaller than the chemical potential in the presence of a magnetic field which is of the order of the square of the chemical potential. This means that the quarks occupy only the lowest landau level. We start with a Lagrangian with $3$ colors and $3$ flavors of quarks interacting via an NJL type four-Fermi interaction in the color-flavor locking channel in the presence of a $U(1)_{\tilde{em}}$ gauge field with a baryon chemical potential $\mu$
\bea
\mathcal{L}=\bar{\psi}(i\slashed{\partial}+e\tilde{Q}\slashed{\tilde{A}}+\mu\gamma_0)\psi +\sum_{\eta=1}^{3}\frac{G}{4}(\bar{\psi}P_{\eta}\psi_c)(\bar{\psi_c}P_{\eta}\psi)
\eea 
where $\psi_{ia}$ is the quark field with $a$ being the flavor index and $i$ being the color index. Also the field $\psi_c$ is the charge conjugate of the field $\psi$ given by $\psi_c=C\bar{\psi}^T$ where $C=i\gamma^2\gamma^0$. $\tilde{A}_{\mu}$ is the $U(1)_{\tilde{em}}$ gauge field. For simplicity we restrict ourselves to the anti-symmetric color-flavor locked channel. Hence, $\left(P_{\eta}\right)_{ia-jb}=i\gamma_5\epsilon_{ij\eta}\epsilon_{ab\eta}$. $\tilde{Q}$ is the rotated $U(1)_{\tilde{em}}$ charge of the quarks and is given by $(+1)\Omega_{+}+(-1)\Omega_{-}+(0)\Omega_0$ where $\Omega_{+}, \Omega_{-}$ and $\Omega_0$ are projectors of positively charged quarks, negatively charged quarks and neutral quarks. The operators $\Omega_{\pm}, \Omega_{0}$ can be expressed as follows in a notation that will be explained in the few lines that follow,
\bea
\Omega_{+}=
\begin{pmatrix}
0 & 0 & 0 & 0 & 0 & 0 & 0 & 0 & 0 \\
 0 & 0 & 0 & 0 & 0 & 0 & 0 & 0 & 0 \\
 0 & 0 & 0 & 0 & 0 & 0 & 0 & 0 & 0 \\
 0 & 0 & 0 & 0 & 0 & 0 & 0 & 0 & 0 \\
 0 & 0 & 0 & 0 & 0 & 0 & 0 & 0 & 0 \\
 0 & 0 & 0 & 0 & 0 & 0 & 0 & 0 & 0 \\
 0 & 0 & 0 & 0 & 0 & 0 & 1 & 0 & 0 \\
 0 & 0 & 0 & 0 & 0 & 0 & 0 & 1 & 0 \\
 0 & 0 & 0 & 0 & 0 & 0 & 0 & 0 & 0 
 \end{pmatrix}, 
 \Omega_{-}=
\begin{pmatrix}
0 & 0 & 0 & 0 & 0 & 0 & 0 & 0 & 0 \\
 0 & 0 & 0 & 0 & 0 & 0 & 0 & 0 & 0 \\
 0 & 0 & 1 & 0 & 0 & 0 & 0 & 0 & 0 \\
 0 & 0 & 0 & 0 & 0 & 0 & 0 & 0 & 0 \\
 0 & 0 & 0 & 0 & 0 & 0 & 0 & 0 & 0 \\
 0 & 0 & 0 & 0 & 0 & 1 & 0 & 0 & 0 \\
 0 & 0 & 0 & 0 & 0 & 0 & 0 & 0 & 0 \\
 0 & 0 & 0 & 0 & 0 & 0 & 0 & 0 & 0 \\
 0 & 0 & 0 & 0 & 0 & 0 & 0 & 0 & 0 
 \end{pmatrix}, 
 \Omega_{0}=
\begin{pmatrix}
1 & 0 & 0 & 0 & 0 & 0 & 0 & 0 & 0 \\
 0 & 1 & 0 & 0 & 0 & 0 & 0 & 0 & 0 \\
 0 & 0 & 0 & 0 & 0 & 0 & 0 & 0 & 0 \\
 0 & 0 & 0 & 1 & 0 & 0 & 0 & 0 & 0 \\
 0 & 0 & 0 & 0 & 1 & 0 & 0 & 0 & 0 \\
 0 & 0 & 0 & 0 & 0 & 0 & 0 & 0 & 0 \\
 0 & 0 & 0 & 0 & 0 & 0 & 0 & 0 & 0 \\
 0 & 0 & 0 & 0 & 0 & 0 & 0 & 0 & 0 \\
 0 & 0 & 0 & 0 & 0 & 0 & 0 & 0 & 1 
 \end{pmatrix}.
 \eea
To clarify our notation, for $i=\pm,0$, $(\Omega_{i})_{ia-jb}$ is written as a $9\times 9$ matrix above, where, $3\times 3$ combinations of 3 flavors($a=1,2,3$) and 3 colors ($i=1,2,3$) are denoted by the $9$ rows of the matrix and $3\times 3$ combinations of 3 flavors($b=1,2,3$) and 3 colors ($j=1,2,3$) are denoted by the $9$ columns of the matrix. $G$ is the coupling constant with dimension of $\text{length}^2$. As we are interested in the bosonic excitations about the condensate, we need to introduce auxiliary fields in the action. Remembering that we are considering pairing only in the anti-symmetric channel in color and flavor, we introduce three auxiliary fields $\Delta_1,\Delta_2$ and $\Delta_3$ and rewrite the action as  
\bea
\mathcal{S}=-\frac{\Delta_1^2 +\Delta_2^2 +\Delta_3^2}{G}+\frac{1}{2}\text{Tr}\left(\text{Log}\left(S_{\text{quark}}^{-1}\right)\right).
\label{action2}
\eea
The matrix $S_{\text{quark}}^{-1}$ is given by
\bea
S_{\text{quark}}^{-1}=
\begin{pmatrix}
i\slashed{\partial}+\tilde{e}\tilde{Q}\slashed{\tilde{A}}+\mu\gamma_0 && i\gamma^5\Delta^- \\
-i(\gamma_0 \gamma^5 (\Delta^-)^{\dagger}\gamma_0) && i\slashed{\partial}+\tilde{e}\tilde{Q}\slashed{\tilde{A}}-\mu\gamma_0
\end{pmatrix}
\label{Squark}
\eea where $\Delta^{-}$ represents the auxiliary fields in our $9\times 9$ matrix notation as
\bea
\Delta^{-} =\begin{pmatrix}
0 & 0 & 0 & 0 & \Delta_3 & 0 & 0 & 0 & \Delta_2\\
0 & 0 & 0 & -\Delta_3 & 0 & 0 & 0 & 0 & 0\\
0 & 0 & 0 & 0 & 0 & 0 & -\Delta_2 & 0 & 0\\
0 & -\Delta_3 & 0 & 0 & 0 & 0 & 0 & 0 & 0\\
\Delta_3 & 0 & 0 & 0 & 0 & 0 & 0 & 0 & \Delta_1\\
0 & 0 & 0 & 0 & 0 & 0 & 0 & -\Delta_1 & 0\\
0 & 0 & -\Delta_2 & 0 & 0 & 0 & 0 & 0 & 0\\
0 & 0 & 0 & 0 & 0 & -\Delta_1 & 0 & 0 & 0\\
\Delta_2 & 0 & 0 & 0 & \Delta_1 & 0 & 0 & 0 & 0
\end{pmatrix}.
\eea
We introduce an external magnetic field at this point in the problem by assigning $\tilde{A}^{\mu}=(0,0,\tilde{B}x,0)$. The idea is to first minimize the action of Eq. \ref{action2} with respect to the magnitude of the auxiliary fields which gives us three gap equations from which we can extract an estimate of the magnitudes of the three gaps. Then we expand the action of Eq. \ref{action2} about this minimum, to quadratic order in the fluctuations to obtain the leading order terms of the effective action for the Goldstone modes. For simplicity we choose to work in a regime where the three gaps are equal, i.e, $\Delta_1=\Delta_2=\Delta_3\equiv\Delta$. But in order to perform a consistent calculation we need to figure out where in the QCD phase diagram ($\mu$ vs $\tilde{B}$) this regime is or at what value of the magnetic field the three gaps equal each other. But before we do that, the form of the condensate can be simplified further by symmetry arguments. Note that as the Lagrangian possesses $SU(2)_{R+L+C}$ global symmetry and so does the condensate, we should have $\Delta_1=\Delta_2$. This is as far as we can go before solving for the gaps explicitly using the gap equations. To solve for the gaps we write down the action as a function of $\Delta_1$ and $\Delta_3$ 
\bea
\mathcal{S}[\Delta_1,\Delta_3]=\frac{8\Delta_1\Delta_1^{\dagger} +4\Delta_3\Delta_3^{\dagger}}{4G}-\frac{1}{2}\text{Tr}(\text{Log}(S_{\text{quark}}^{-1})).
\label{action3}
\eea
Minimizing the action of Eq. \ref{action3} with respect to $\Delta_1$ and $\Delta_3$ we obtain the gap equations for the two gaps. The gap equation for $\Delta_1$ is given by 
\bea
\frac{d\mathcal{S}}{d\Delta_{1}^{\dagger}}\bigg|_{\Delta_1=\Delta_3=\Delta}=\frac{8\Delta_{1}}{4G}\bigg|_{\Delta_1=\Delta_3=\Delta}-\frac{1}{2}\text{Tr}\left(S_{\text{quark}}.\frac{dS_{\text{quark}}^{-1}}{d\Delta_{1}^{\dagger}}\right)\bigg|_{\Delta_1=\Delta_3=\Delta}=0
\label{d1}
\eea
and for $\Delta_3$ it is 
\bea
\frac{d\mathcal{S}}{d\Delta_{3}^{\dagger}}\bigg|_{\Delta_1=\Delta_3=\Delta}=\frac{4\Delta_{3}}{4}\bigg|_{\Delta_1=\Delta_3=\Delta}-\frac{1}{2}\text{Tr}\left(S_{\text{quark}}.\frac{dS_{\text{quark}}^{-1}}{d\Delta_{3}^{\dagger}}\right)\bigg|_{\Delta_1=\Delta_3=\Delta}=0.
\label{d3}
\eea. In order to proceed further we need to evaluate $S_{\text{quark}}$ in momentum space as the calculations are much simpler in it than in position space. However, as is evident from the position space form of $S_{\text{quark}}^{-1}$ given in Eq. \ref{Squark}, the inverse propagator appears to not be translationally invariant for the charged quarks, which means that it is not diagonal in the momentum space and we lose the advantage we had in going to the momentum space. However, this can be remedied by using the Landau level basis for the quarks instead of the standard free particle basis. As the eigen states of a fermion in a magnetic field is given by the Landau levels, it is natural to expect the propagator to be diagonal in this basis. This method was first introduced by Ritus in the year $1972$ \cite{Ritus:1972ky}. We write down the explicit form of the quark propagator in the Landau level basis, where $l$ is the Landau level index, in the appendix in order to avoid interrupting the thread of logic here. Note that in the appendix we derived the quark propagator only in the limit $\Delta_1=\Delta_2=\Delta_3$. However, this will suffice for the purpose of solving Eq. \ref{d1} and Eq. \ref{d3} as both of them are to be solved for $\Delta_1=\Delta_2=\Delta_3$. Also we will be working in the limit of large magnetic fields, of the order of the square of the chemical potential so that only the lowest Landau level is occupied. As mentioned before, we simplify \ref{d1} and \ref{d3} using the expression for the propagator in the appendix and obtain two equations for $\Delta$. By demanding the two be consistent with each other we can determine the magnetic field as a function of the chemical potential for which $\Delta_1=\Delta_3$. Simplifying \ref{d1} we get 
\bea
\frac{2}{g^2}=\frac{2\mu^2}{\pi^2}\text{log}\left(\frac{2\Lambda}{\Delta}\right)+\frac{\tilde{e}\tilde{B}}{\pi^2}\text{Log}\left(\frac{2\Lambda}{\Delta}\right)
\label{d11}
\eea
and simplifying \ref{d3} gives 
\bea
\frac{1}{g^2}=2\frac{\mu^2}{\pi^2}\text{log}\left(\frac{2\Lambda}{\Delta}\right)
\label{d33}
\eea. $\Lambda$ is the cut-off used for the divergent integrals in the gap equation. It is worthwhile mentioning at this point that although we introduce a sharp ultraviolet cutoff here for the sake of clarity, the cancellation of the UV divergences in the calculations that follow do not depend on any kind of regularization procedure. \ref{d11} and \ref{d33} can only be consistent when $\tilde{e}\tilde{B}=2\mu^2$. This brings us to the reason why we are considering $\tilde{e}\tilde{B}\sim 2\mu^2$ in the first place and it is because we need both high magnetic fileds $\tilde{e}\tilde{B}\geq \frac{\mu^2}{2}$ and all the gaps to be equal to simplify our calculations. The motivation is to obtain analytic results in a regime where the effect of the magnetic field on the Goldstone mode propagators would be considerable. Let us introduce a matrix at this point given by $\bar{\Delta}\equiv \Delta^-\big|_{\Delta_1=\Delta_3=\Delta}$.
Having obtained the strength of the magnetic field for which the gaps are equal let us now try to find the propagator for the neutral Goldstone modes. Before we do that we write them out explicitly here for clarity. Note that the gap transforms under chiral flavor rotation as 
\begin{align}
\begin{split}
\Delta' &= e^{i\gamma^5G^a\alpha^a}\bar{\Delta} e^{i\gamma^5(G^a)^T\alpha^a}\\
&= \left(1+i\gamma^5G^a\alpha^a\right)\bar{\Delta}\left(1+i\gamma^5(G^a)^T\alpha^a\right)+..\\
&= \bar{\Delta} + i\alpha^a \gamma^5 M^a +...,   \text{          for $a=1,..,8$}
\end{split}
\end{align}
where, $G^a$ are generators of $SU(3)$ gauge group, $\alpha^a$ are fields corresponding to the generators $G^a$ and $M^a=G^a\bar{\Delta}+\bar{\Delta}(G^a)^T$. Similarly, under an axial $U(1)$ rotation which happens to be an approximate symmetry of the Lagrangian at high density, the condensate transforms as
\begin{align}
\begin{split}
\Delta' &= e^{i\gamma^5\alpha^9}\bar{\Delta} e^{i\gamma^5\alpha^9}\\
&= \bar{\Delta} +2i\alpha^{9}\gamma^5\bar{\Delta}+...\equiv \bar{\Delta}+ i\alpha^{9}\gamma^5M^9 +..
\end{split}
\end{align}
and under an ordinary $U(1)$ rotation
\begin{align}
\begin{split}
\Delta' &= e^{i\alpha^{10}}\bar{\Delta} e^{i\alpha^{10}}\\
&= \bar{\Delta} +2i\alpha^{10}\bar{\Delta}+..\equiv \bar{\Delta}+ i\alpha^{10}M^{10}+...
\end{split}
\end{align}
Among the $10$ modes $M^a$ we need to find the neutral ones. The neutral goldstone modes are the ones for which $\tilde{Q}.M^a+M^a.\tilde{Q}=0$. We find that $M^1, M^2, M^3$, $M^8$, $M^9$ and $M^{10}$ are neutral as expected and the rest are not. We describe the computation of the propagator for $\alpha^3$ in some detail here and state the result for the rest of the five as the calculations are very similar. To proceed we first need to rewrite the action of Eq. \ref{action2} in terms of $\Delta^-$ as
\bea
\mathcal{S}=-\frac{1}{4G}\Delta^-_{ia-jb}(\Delta^-)^\dagger_{jb-ia}+\frac{1}{2}\text{Tr}\left(\text{Log}\left(S_{quark}^{-1}\right)\right).
\eea
Now we expand the above action about $\Delta^-=\bar{\Delta}$ upto second order in fluctuations($\delta\Delta$) to obtain the following expression
\begin{align}
\begin{split}
\mathcal{S}[\Delta]&= \mathcal{S}[\bar{\Delta}]+\frac{\partial\mathcal{S}}{\partial\Delta^-_{ia-jb}}\bigg|_{\Delta^-=\bar{\Delta}}\delta\Delta_{ia-jb} +\frac{\partial\mathcal{S}}{\partial(\Delta^-)^\dagger}_{ia-jb}\bigg|_{\Delta^-=\bar{\Delta}}\delta\Delta^{\dagger}_{ia-jb}+ \frac{1}{2}\frac{\partial^2\mathcal{S}}{\partial\Delta^-_{kd-lf}\partial\Delta^-_{ia-jb}}\bigg|_{\Delta^-=\bar{\Delta}}\delta\Delta_{kd-lf}\delta\Delta_{ia-jb}+\\
&\qquad\quad \frac{1}{2}\frac{\partial^2\mathcal{S}}{\partial(\Delta^-)^\dagger_{kd-lf}\partial(\Delta^-)^\dagger_{ia-jb}}\bigg|_{\Delta^-=\bar{\Delta}}\delta\Delta^\dagger_{kd-lf}\delta\Delta^\dagger_{ia-jb}+ \frac{1}{2}\frac{\partial^2\mathcal{S}}{\partial(\Delta^-)^\dagger_{kd-lf}\partial\Delta^-_{ia-jb}}\bigg|_{\Delta^-=\bar{\Delta}}\delta\Delta^\dagger_{kd-lf}\delta\Delta_{ia-jb}+\\
&\qquad\quad \frac{1}{2}\frac{\partial^2\mathcal{S}}{\partial\Delta^-_{kd-lf}\partial(\Delta^-)^\dagger_{ia-jb}}\bigg|_{\Delta^-=\bar{\Delta}}\delta\Delta_{kd-lf}\delta\Delta^\dagger_{ia-jb}.
\end{split}\label{expand}\end{align} 
The coefficients of the terms in the above expression which are first order in fluctuations $\delta\Delta$ go to zero due to the gap equation. Also note that there is an easy trick to simplify second derivatives of the trace of logarithm of the inverse quark propagator with respect to the gap matrix and it is the following,
\bea
\frac{\partial^2 \text{Tr}\left(\text{Log}(S_{\text{quark}}^{-1})\right)}{\partial\Delta^-_{ia-jb}\partial\Delta^-_{kd-lf}}=- \text{Tr}\left(S_{\text{quark}}\frac{dS_{\text{quark}}^{-1}}{\partial\Delta^-_{ia-jb}}S_{\text{quark}}\frac{dS_{\text{quark}}^{-1}}{\partial\Delta^-_{kd-lf}}\right).
\label{trick}
\eea   
Although we have explicitly written out the above expression for the second derivative with respect to $\Delta^-$, similar relations are true for second derivatives with respect to $(\Delta^-)^{\dagger}$ and mixed second derivatives with respect to $\Delta^-$ and $(\Delta^-)^{\dagger}$. Now we set $\delta\Delta=i\alpha^3\gamma^5 M^3$ in Eq. \ref{expand} as we are interested in the propagator of $\alpha^3$ and denote the part of the action that is quadratic in $\alpha^3$ as $\mathcal{S}_{3}$. After using Eq. \ref{trick} and the expressions for the quark propagators for the $0^{th}$ Landau level from the appendix and some tedious algebra we finally get 
\begin{align}
\begin{split}
\mathcal{S}_{3}[\Delta]&= \mathcal{S}[\bar{\Delta}]-\frac{1}{4G}3\Delta^2(\alpha^3)^2\\
&\qquad\quad -\frac{\Delta^2(\alpha^3)^2}{4}\left[\int\frac{d^4p}{(2\pi)^4}\left(1+\frac{\mathbf{p}.(\mathbf{p}-\mathbf{k})}{|\mathbf{p}||\mathbf{p}-\mathbf{k}|}\right)\frac{4(p_0(p_0-k_0)-4(|\mathbf{p}-\mathbf{k}|-\mu)(|\mathbf{p}|-\mu)-4\Delta^2)}{\left((p_0-k_0)^2-(|\mathbf{p}-\mathbf{k}|-\mu)^2-\Delta^2\right)\left(p_0^2-(|\mathbf{p}|-\mu)^2-\Delta^2\right)}\right]\\
&\qquad\quad -\frac{\Delta^2(\alpha^3)^2}{4}\left[\int \frac{dp_0dp_3}{(2\pi)^4}\frac{1}{2}\frac{4\pi^2\tilde{e}\tilde{B}}{4}\frac{16}{2\pi}\frac{p_0(p_0-k_0)-(|p_3-k_3|-\mu)(|p_3|-\mu)-\Delta^2}{\left((p_0-k_0)^2-(|p_3-k_3|-\mu)^2-\Delta^2\right)\left(p_0^2-(|p_3|-\mu)^2-\Delta^2\right)}+\mathcal{O}\left(\frac{k_{\perp}^2}{\tilde{e}\tilde{B}}\right)\right]+...
\end{split}\label{expand1}\end{align} 

In Eq. \ref{expand1} we first do the $p_0$ integral followed by the $\mathbf{p}$ and $p_3$ integrals. The three momentum integrals and the $p_3$ integrals are performed about the Fermi surface, $|\mathbf{p}|\sim \mu$ and $|p_3|\sim\mu$. Note that Eq. \ref{expand1} has integrals which are divergent and we need to carefully subtract these divergences to obtain our final result. The divergences basically cancell with the term with the coupling constant ($G$) in the denominator in Eq. \ref{expand1}. To see how exactly this cancellation works we reorganize Eq. \ref{expand1} in a form that is more convenient
\begin{align}
\begin{split}
\mathcal{S}_{3}[\Delta]&= \mathcal{S}[\bar{\Delta}]-\frac{1}{4G}3\Delta^2(\alpha^3)^2-\int\frac{d^4p}{(2\pi)^4}\frac{2}{(p_0-k_0)^2-(|\mathbf{p}|-\mu)^2-\Delta^2}-\frac{|\tilde{e}\tilde{B}|}{4\pi}\int\frac{dp_0dp_3}{(2\pi)^2}\frac{1}{2}\frac{2}{(p_0-k_0)^2-(|p_3|-\mu)^2-\Delta^2}\\
&\quad\quad +\frac{\Delta^2(\alpha^3)^2}{4}\left[\int\frac{d^4p}{(2\pi)^4}\frac{8p_0k_0}{\left((p_0-k_0)^2-(|\mathbf{p}|-\mu)^2-\Delta^2\right)\left(p_0^2-(|\mathbf{p}|-\mu)^2-\Delta^2\right)}\right]\\
&\qquad\quad +\frac{\Delta^2(\alpha^3)^2}{4}\left[\int \frac{dp_0dp_3}{(2\pi)^4}\frac{1}{2}\frac{4\pi^2\tilde{e}\tilde{B}}{4}\frac{16}{2\pi}\frac{p_0k_0}{\left((p_0-k_0)^2-(|p_3|-\mu)^2-\Delta^2\right)\left(p_0^2-(|p_3|-\mu)^2-\Delta^2\right)}  \right]\\
&\qquad\quad -\frac{\Delta^2(\alpha^3)^2}{4}\left[\int\frac{d^4p}{(2\pi)^4}\left(1+\frac{\mathbf{p}.(\mathbf{p}-\mathbf{k})}{|\mathbf{p}||\mathbf{p}-\mathbf{k}|}\right)\frac{4(p_0(p_0-k_0)-4(|\mathbf{p}-\mathbf{k}|-\mu)(|\mathbf{p}|-\mu)-4\Delta^2)}{\left((p_0-k_0)^2-(|\mathbf{p}-\mathbf{k}|-\mu)^2-\Delta^2\right)\left(p_0^2-(|\mathbf{p}|-\mu)^2-\Delta^2\right)}\right.\\
&\qquad\quad \left.-\int\frac{d^4p}{(2\pi)^4}\frac{8p_0(p_0-k_0)-8(|\mathbf{p}|-\mu)^2-8\Delta^2}{\left((p_0-k_0)^2-(|\mathbf{p}|-\mu)^2-\Delta^2\right)\left(p_0^2-(|\mathbf{p}|-\mu)^2-\Delta^2\right)}\right]\\
&\qquad\quad -\frac{\Delta^2(\alpha^3)^2}{4}\left[\int \frac{dp_0dp_3}{(2\pi)^4}\frac{1}{2}\frac{4\pi^2\tilde{e}\tilde{B}}{4}\frac{16}{2\pi}\frac{p_0(p_0-k_0)-(|p_3-k_3|-\mu)(|p_3|-\mu)-\Delta^2}{\left((p_0-k_0)^2-(|p_3-k_3|-\mu)^2-\Delta^2\right)\left(p_0^2-(|p_3|-\mu)^2-\Delta^2\right)}\right.\\
&\qquad\quad \left.-\int \frac{dp_0dp_3}{(2\pi)^4}\frac{1}{2}\frac{4\pi^2\tilde{e}\tilde{B}}{4}\frac{16}{2\pi}\frac{p_0(p_0-k_0)-(|p_3|-\mu)^2-\Delta^2}{\left((p_0-k_0)^2-(|p_3|-\mu)^2-\Delta^2\right)\left(p_0^2-(|p_3|-\mu)^2-\Delta^2\right)}  \right]+...
\end{split}\label{expand2}\end{align}. The third and the fourth term in Eq. \ref{expand2} contain the leading divergences in the expansion $\frac{|\mathbf{k}|}{\mu}$. Now we need to use the gap equation to cancell these divergent integrals with the second term in Eq. \ref{expand2}. Using the gap equation at $\tilde{e}\tilde{B} \sim 2\mu^2$ we can see that 
\bea
\frac{3}{4G}=-i\left(\int\frac{d^4p}{(2\pi)^4}\frac{2}{(p_0)^2-(|\mathbf{p}|-\mu)^2-\Delta^2}+\frac{|\tilde{e}\tilde{B}|}{4\pi}\int\frac{dp_0dp_3}{(2\pi)^2}\frac{1}{2}\frac{2}{(p_0)^2-(|p_3|-\mu)^2-\Delta^2}\right)
\label{g}
\eea
which achieves the cancellation we were looking for and we are left with an action for $\alpha^3$ given by
\begin{align}
\begin{split}
\mathcal{S}_{3}[\Delta]&= \mathcal{S}[\bar{\Delta}]+\Delta^2(\alpha^3)^2\int\frac{d^4p}{(2\pi)^4}\frac{k_0^2}{(p_0^2-(|\mathbf{p}|-\mu)^2-\Delta^2)((p_0-k_0)^2-(|\mathbf{p}|-\mu)^2-\Delta^2)}\\
&\qquad\quad +\Delta^2(\alpha^3)^2\int\frac{dp_0dp_3}{(2\pi)^2}\frac{1}{2}\frac{|\tilde{e}\tilde{B}|}{4\pi}\frac{k_0^2}{(p_0^2-(|p_3|-\mu)^2-\Delta^2)((p_0-k_0)^2-(|p_3|-\mu)^2-\Delta^2)}\\
&\qquad\quad -\frac{\Delta^2(\alpha^3)^2}{4}\int\frac{d^4p}{(2\pi)^4}\left[\left(1+\frac{\mathbf{p}.(\mathbf{p}-\mathbf{k})}{|\mathbf{p}||\mathbf{p}-\mathbf{k}|}\right)\frac{4(p_0(p_0-k_0)-4(|\mathbf{p}-\mathbf{k}|-\mu)(|\mathbf{p}|-\mu)-4\Delta^2)}{\left((p_0-k_0)^2-(|\mathbf{p}-\mathbf{k}|-\mu)^2-\Delta^2\right)\left(p_0^2-(|\mathbf{p}|-\mu)^2-\Delta^2\right)}\right.\\
&\qquad\quad\left. -\left(1+\frac{\mathbf{p}.(\mathbf{p}-\mathbf{k})}{|\mathbf{p}||\mathbf{p}-\mathbf{k}|}\right)\frac{4(p_0(p_0-k_0)-4(|\mathbf{p}|-\mu)(|\mathbf{p}|-\mu)-4\Delta^2)}{\left((p_0-k_0)^2-(|\mathbf{p}|-\mu)^2-\Delta^2\right)\left(p_0^2-(|\mathbf{p}|-\mu)^2-\Delta^2\right)} \right]\\
&\qquad\quad -\frac{\Delta^2(\alpha^3)^2}{4}\int\frac{d^4p}{(2\pi)^4}\left(\frac{\mathbf{p}.(\mathbf{p}-\mathbf{k})}{|\mathbf{p}||\mathbf{p}-\mathbf{k}|}-1\right)\frac{4(p_0(p_0-k_0)-4(|\mathbf{p}|-\mu)^2-4\Delta^2)}{\left((p_0-k_0)^2-(|\mathbf{p}|-\mu)^2-\Delta^2\right)\left(p_0^2-(|\mathbf{p}|-\mu)^2-\Delta^2\right)}\\
&\qquad\quad -\Delta^2(\alpha^3)^2\int\frac{dp_0dp_3}{2(2\pi)^2}\frac{1}{2}\frac{|\tilde{e}\tilde{B}|}{2\pi}\left[\frac{2p_0(p_0-k_0)-2(|p_3-k_3|-\mu)(|p_3|-\mu)-2\Delta^2}{((p_0-k_0)^2-(|p_3-k_3|-\mu)^2-\Delta^2)(p_0^2-(|p_3|-\mu)^2-\Delta^2)}\right.\\
&\qquad\quad\left. -\frac{2p_0(p_0-k_0)-2(|p_3|-\mu)^2-2\Delta^2}{((p_0-k_0)^2-(|p_3|-\mu)^2-\Delta^2)(p_0^2-(|p_3|-\mu)^2-\Delta^2)}\right]+...
\end{split}\label{expand3}\end{align}. 
The above action has only finite integrals except for one term with a factor of $\left(\frac{\mathbf{p}.(\mathbf{p}-\mathbf{k})}{|\mathbf{p}||\mathbf{p}-\mathbf{k}|}-1\right)$ inside the four momentum integral. Although this term is divergent, it is suppressed by powers of $\frac{|\mathbf{k}|}{\mu}$ and can be ignored for our purpose. If we now expand our result in $\frac{k_0^2}{\Delta^2}$, $\frac{k_3^2}{\Delta^2}$ and $\frac{|\mathbf{k}|^2}{\Delta^2}$ we obtain the leading order quadratic action in $\alpha^3$ as
\begin{align}
\begin{split}
\mathcal{S}_{3}[\Delta] &= \mathcal{S}[\bar{\Delta}]-\frac{i((\alpha^3)^2)\mu^2}{4\pi^2}\left(k_0^2-\frac{|\mathbf{k}|^2}{3}\right)\\
&\qquad\quad -\frac{i((\alpha^3)^2)|\tilde{e}\tilde{B}|}{16\pi^2}(k_0^2-k_3^2)+...
\end{split}
\label{disp}
\end{align}
We can perform similar calculations for the other neutral Goldstone modes $1, 2, 8, 9, 10$ and find the corresponding inverse propagators. 
From these expressions we can extract the speed of the six neutral Goldstone modes. If we denote the speed of the $i$ th excitaion perpendicular and parallel to the magnetic field as $v^i_{\perp}$ and $v^i_{\parallel}$ respectively, we have $(v^i)_{\perp}^2=\frac{2}{9}, (v^i)_{\parallel}^2=\frac{4}{9}$ for $i=1, 2, 3$, $(v^i)_{\perp}^2=\frac{8}{27}, (v^i)_{\parallel}^2=\frac{11}{27}$ for $i=8$ and $(v^i)_{\perp}^2=\frac{5}{27}, (v^i)_{\parallel}^2=\frac{17}{27}$ for $i=9, 10$ for the values of the magnetic field for which $\tilde{e}\tilde{B}\sim 2\mu^2$. 

\section{Lower Magnetic Fields}\label{magD}

The calculation in the section above although applicable to only $\tilde{e}\tilde{B}\sim 2\mu^2$ paves the way for a similar calculation in this section for more realistic magnetic fields. In this section we discuss the propagators of the neutral Goldstone modes for magnetic fields $\tilde{e}\tilde{B}\leq 0.3 \mu^2$. Just to clarify the separation of scales in this discussion, if the momentum scale at which we are looking at the propagators of the Goldstone modes is denoted by $k$, we will be concentrating in a regime $|\mathbf{k}|^2< \Delta^2\leq \tilde{e}\tilde{B}< 0.3 \mu^2$. This regime is such that multiple Landau levels are filled and the total number of Landau levels occupied by the quarks is given by $\left[\frac{\mu^2}{2\tilde{e}\tilde{B}}\right]$, where $\left[..\right]$ denotes the greatest integer smaller than $\frac{\mu^2}{2\tilde{e}\tilde{B}}$. Before we delve into the details of the effective theory in this regime, it should be noted that the effect of multiple Landau levels on the gap equation in the presence of a magnetic field was discussed in \cite{PhysRevD.76.105030}. It was found that the two gaps $\Delta_1$ and $\Delta_3$ exhibit de Haas-van Alphen oscillations. This is a consequence of the fact that the number of occupied Landau levels changes discontinuously as the magnetic field is varied with respect to the chemical potential. From \cite{PhysRevD.76.105030} we can see that the two gaps $\Delta_1$ and $\Delta_3$ become appreciably different from each other for $\tilde{e}\tilde{B}\geq 0.5\mu^2$. Note that as the magnetic field is increased further, the two gaps become equal to each other at $\tilde{e}\tilde{B}\sim 2\mu^2$ which is consistent with our results in the previous section. The effect of the de Haas-van Alphen oscillation is not very significant for magnetic fields $\tilde{e}\tilde{B}<0.3\mu^2$ and it is reasonable to use $\Delta_1=\Delta_3$ in this regime. As we lower the magnetic field further this approximation becomes more accurate\cite{PhysRevD.76.105030}. \\
\begin{figure}

\begin{tabular}{ccc}
\subfloat[]
{\label{fig:f0}
\includegraphics[width=0.4\textwidth]{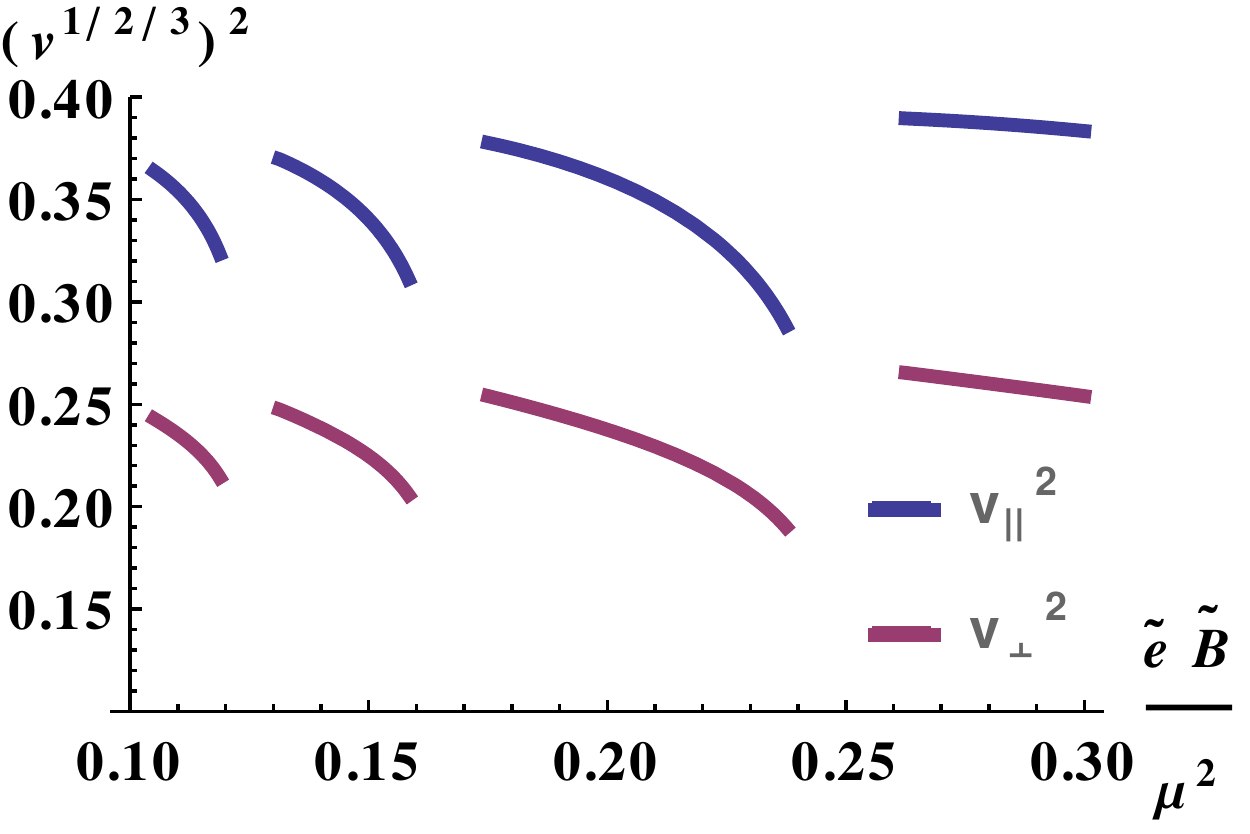}}

\subfloat[]
{\label{fig:f1}
\includegraphics[width=0.4\textwidth]{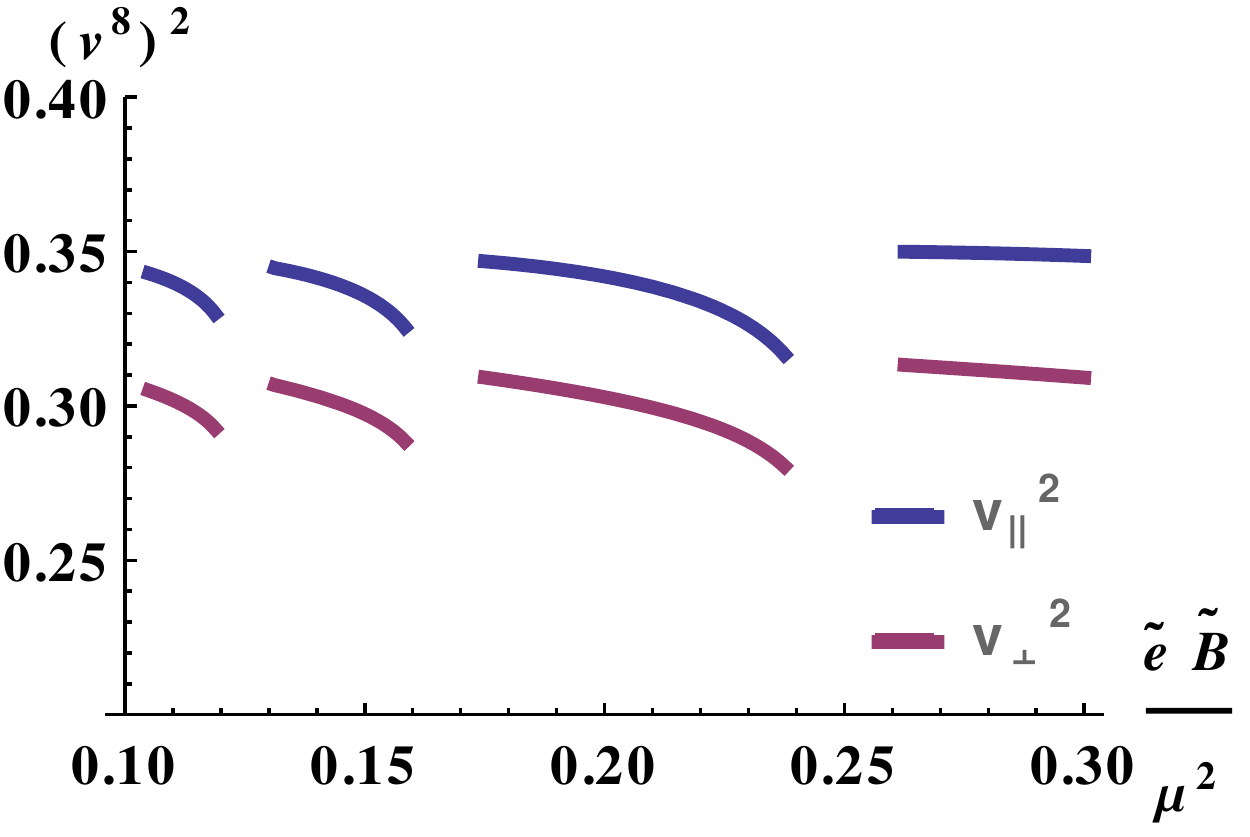}}\\
\subfloat[]
{\label{fig:f2}
\includegraphics[width=0.4\textwidth]{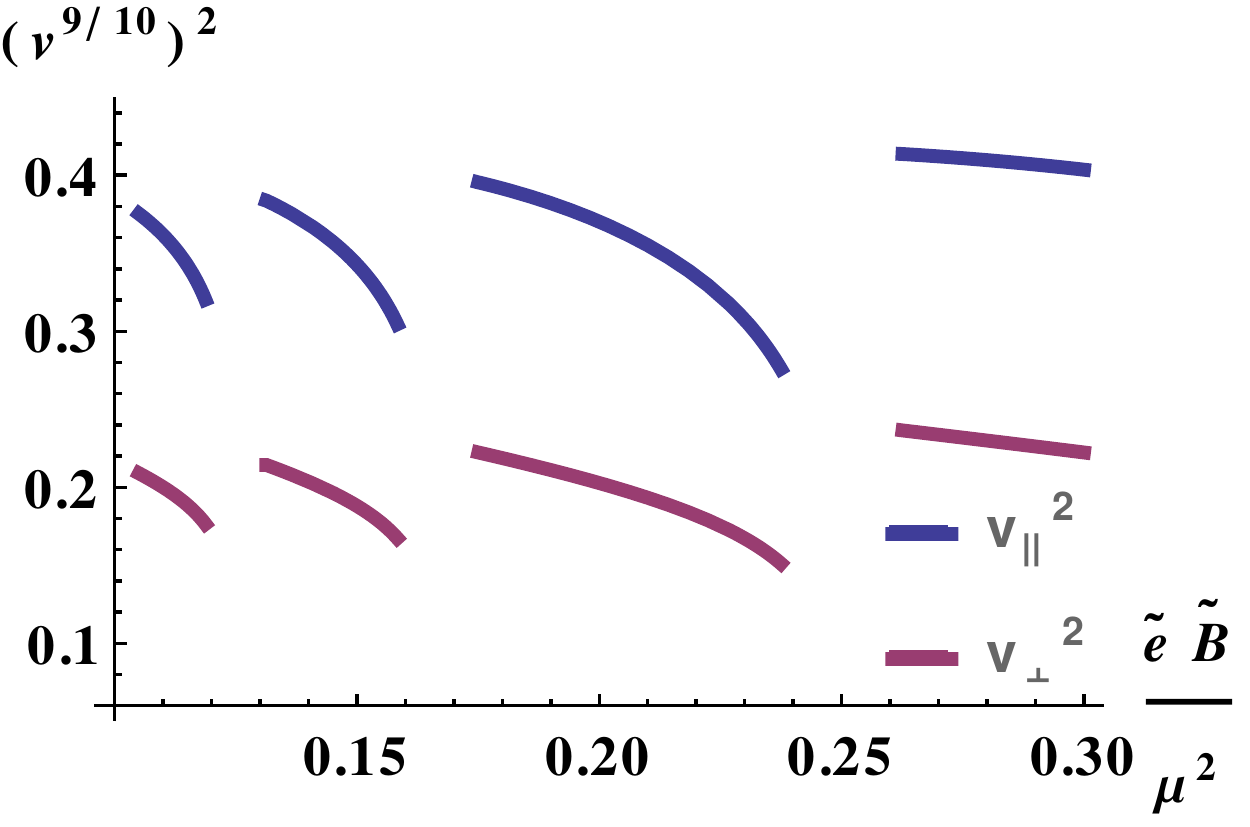}}
\end{tabular}
\caption{\label{fig:multi_panel} 
Fig. \ref{fig:f0} shows speed in directions parallel and perpendicular to the magnetic fields for the mesonic modes $1, 2, 3$ as a function of $\tilde{e}\tilde{B}$ where as Fig. \ref{fig:f1} shows speed in directions parallel and perpendicular to the magnetic fields for the $8^{th}$ mesonic mode as a function of $\tilde{e}\tilde{B}$. Fig. \ref{fig:f2} is a similar plot showing the speeds of mode $9$ and $10$. }\label{fig:speed}
\end{figure} 

Although it is probably difficult to treat this problem in a regime where multiple Landau levels have to be taken into account analytically in its full generality, it is possible to do so in the limit $|\mathbf{k}|\ll\Delta<\tilde{e}\tilde{B}$, if we were to consider only leading order terms in an expansion in $\frac{|\mathbf{k}|^2}{\tilde{e}\tilde{B}}$. In this limit we can obtain an expression for the effective action for the Goldstone modes analogous to \ref{expand1}. The only difference is that in addition to a term like $-\frac{\Delta^2(\alpha^3)^2}{4}\left[\int \frac{dp_0dp_3}{(2\pi)^4}\frac{1}{2}\frac{4\pi^2\tilde{e}\tilde{B}}{4}\frac{16}{2\pi}\frac{p_0(p_0-k_0)-(|p_3-k_3|-\mu)(|p_3|-\mu)-\Delta^2}{\left((p_0-k_0)^2-(|p_3-k_3|-\mu)^2-\Delta^2\right)\left(p_0^2-(|p_3|-\mu)^2-\Delta^2\right)}+\mathcal{O}\left(\frac{k_{\perp}^2}{\tilde{e}\tilde{B}}\right)\right]$ as in \ref{expand1}, we also have $-\frac{\Delta^2(\alpha^3)^2}{4}\left[\int \frac{dp_0dp_3}{(2\pi)^4}\frac{4\pi^2\tilde{e}\tilde{B}}{4}\frac{16}{2\pi}\sum_{l=1}^{l_{max}}\frac{p_0(p_0-k_0)-\left(\sqrt{|p_3-k_3|^2+2|\tilde{e}\tilde{B}|l}-\mu\right)\left(\sqrt{|p_3|^2+2\tilde{e}\tilde{B}l}-\mu\right)-\Delta^2}{\left(\left(p_0-k_0\right)^2-\left(\sqrt{|p_3-k_3|^2+2\tilde{e}\tilde{B}l}-\mu\right)^2-\Delta^2\right)\left(p_0^2-\left(\sqrt{|p_3|^2+2\tilde{e}\tilde{B}l}-\mu\right)^2-\Delta^2\right)}+\mathcal{O}\left(\frac{k_{\perp}^2}{\tilde{e}\tilde{B}}\right)\right]$ where $l$ is the Landau level index. The $p_3$ integrals in this additional term which corresponds to $l\neq 0$, are to be performed around $\sqrt{p_3^2+2\tilde{e}\tilde{B}l}\sim\mu$ instead of $|p_3|\sim\mu$. After subtracting the divergences as before and computing the integrals and the sums, we end up with expressions for the speeds of the Goldstone modes along the magnetic field and perpendicular to the field.
Here we state the results only as the manipulations involved are very similar to the ones described in the previous section. The speed of the three modes $\alpha^1,\alpha^2, \alpha^3$ parallel to the magnetic field is given by 
\bea
v_{\parallel}^2(\tilde{e}\tilde{B})=\frac{\frac{\mu^2}{12}+\frac{|\tilde{e}\tilde{B}|}{16}+\frac{|\tilde{e}\tilde{B}|}{16}\frac{2\sqrt{2|\tilde{e}\tilde{B}|}}{\mu}Re\left[i\left[H_{\zeta}\left[-\frac{1}{2},1-\frac{\mu^2}{2|\tilde{e}\tilde{B}|}+\left[\frac{\mu^2}{2|\tilde{e}\tilde{B}|}\right]\right]-H_{\zeta}\left[-\frac{1}{2},1-\frac{\mu^2}{2|\tilde{e}\tilde{B}|}\right]\right]\right]}{\frac{\mu^2}{4}+\frac{|\tilde{e}\tilde{B}|}{16}+\frac{|\tilde{e}\tilde{B}|}{16}\frac{2\mu}{\sqrt{2|\tilde{e}\tilde{B}|}}Re\left[(-i)\left[H_{\zeta}\left[\frac{1}{2},1-\frac{\mu^2}{2|\tilde{e}\tilde{B}|}+\left[\frac{\mu^2}{2|\tilde{e}\tilde{B}|}\right]\right]-H_{\zeta}\left[\frac{1}{2},1-\frac{\mu^2}{2|\tilde{e}\tilde{B}|}\right]\right]\right]}
\eea 
and \bea
v_{\perp}^2(\tilde{e}\tilde{B})=\frac{\frac{\mu^2}{12}}{\frac{\mu^2}{4}+\frac{|\tilde{e}\tilde{B}|}{16}+\frac{|\tilde{e}\tilde{B}|}{16}\frac{2\mu}{\sqrt{2|\tilde{e}\tilde{B}|}}Re\left[(-i)\left[H_{\zeta}\left[\frac{1}{2},1-\frac{\mu^2}{2|\tilde{e}\tilde{B}|}+\left[\frac{\mu^2}{2|\tilde{e}\tilde{B}|}\right]\right]-H_{\zeta}\left[\frac{1}{2},1-\frac{\mu^2}{2|\tilde{e}\tilde{B}|}\right]\right]\right]}
\eea
where $H_{\zeta}$ is the Hurwitz zeta function. 
Similarly for $\alpha^8$ we have 
\bea
v_{\parallel}^2(\tilde{e}\tilde{B})=\frac{\frac{\mu^2}{9}+\frac{|\tilde{e}\tilde{B}|}{48}+\frac{|\tilde{e}\tilde{B}|}{48}\frac{2\sqrt{2|\tilde{e}\tilde{B}|}}{\mu}Re\left[i\left[H_{\zeta}\left[-\frac{1}{2},1-\frac{\mu^2}{2|\tilde{e}\tilde{B}|}+\left[\frac{\mu^2}{2|\tilde{e}\tilde{B}|}\right]\right]-H_{\zeta}\left[-\frac{1}{2},1-\frac{\mu^2}{2|\tilde{e}\tilde{B}|}\right]\right]\right]}{\frac{\mu^2}{3}+\frac{|\tilde{e}\tilde{B}|}{48}+\frac{|\tilde{e}\tilde{B}|}{48}\frac{2\mu}{\sqrt{2|\tilde{e}\tilde{B}|}}Re\left[(-i)\left[H_{\zeta}\left[\frac{1}{2},1-\frac{\mu^2}{2|\tilde{e}\tilde{B}|}+\left[\frac{\mu^2}{2|\tilde{e}\tilde{B}|}\right]\right]-H_{\zeta}\left[\frac{1}{2},1-\frac{\mu^2}{2|\tilde{e}\tilde{B}|}\right]\right]\right]}
\eea 
and \bea
v_{\perp}^2(\tilde{e}\tilde{B})=\frac{\frac{\mu^2}{9}}{\frac{\mu^2}{3}+\frac{|\tilde{e}\tilde{B}|}{48}+\frac{|\tilde{e}\tilde{B}|}{48}\frac{2\mu}{\sqrt{2|\tilde{e}\tilde{B}|}}Re\left[(-i)\left[H_{\zeta}\left[\frac{1}{2},1-\frac{\mu^2}{2|\tilde{e}\tilde{B}|}+\left[\frac{\mu^2}{2|\tilde{e}\tilde{B}|}\right]\right]-H_{\zeta}\left[\frac{1}{2},1-\frac{\mu^2}{2|\tilde{e}\tilde{B}|}\right]\right]\right]}
\eea and for $\alpha^9$ and $\alpha^{10}$
\bea
v_{\parallel}^2(\tilde{e}\tilde{B})=\frac{\frac{5\mu^2}{18}+\frac{|\tilde{e}\tilde{B}|}{3}+\frac{|\tilde{e}\tilde{B}|}{3}\frac{2\sqrt{2|\tilde{e}\tilde{B}|}}{\mu}Re\left[i\left[H_{\zeta}\left[-\frac{1}{2},1-\frac{\mu^2}{2|\tilde{e}\tilde{B}|}+\left[\frac{\mu^2}{2|\tilde{e}\tilde{B}|}\right]\right]-H_{\zeta}\left[-\frac{1}{2},1-\frac{\mu^2}{2|\tilde{e}\tilde{B}|}\right]\right]\right]}{\frac{5\mu^2}{6}+\frac{|\tilde{e}\tilde{B}|}{3}+\frac{|\tilde{e}\tilde{B}|}{3}\frac{2\mu}{\sqrt{2|\tilde{e}\tilde{B}|}}Re\left[(-i)\left[H_{\zeta}\left[\frac{1}{2},1-\frac{\mu^2}{2|\tilde{e}\tilde{B}|}+\left[\frac{\mu^2}{2|\tilde{e}\tilde{B}|}\right]\right]-H_{\zeta}\left[\frac{1}{2},1-\frac{\mu^2}{2|\tilde{e}\tilde{B}|}\right]\right]\right]}
\eea 
and \bea
v_{\perp}^2(\tilde{e}\tilde{B})=\frac{\frac{5\mu^2}{18}}{\frac{5\mu^2}{6}+\frac{|\tilde{e}\tilde{B}|}{3}+\frac{|\tilde{e}\tilde{B}|}{3}\frac{2\mu}{\sqrt{2|\tilde{e}\tilde{B}|}}Re\left[(-i)\left[H_{\zeta}\left[\frac{1}{2},1-\frac{\mu^2}{2|\tilde{e}\tilde{B}|}+\left[\frac{\mu^2}{2|\tilde{e}\tilde{B}|}\right]\right]-H_{\zeta}\left[\frac{1}{2},1-\frac{\mu^2}{2|\tilde{e}\tilde{B}|}\right]\right]\right]}.
\eea
Although the above expresseions are valid for most of $|\mathbf{k}|<\tilde{e}\tilde{B}<0.3\mu^2$, the regime of validity of the above expressions is restricted by $\left(\mu^2-\left[\frac{\mu^2}{2\tilde{e}\tilde{B}}\right]2\tilde{e}\tilde{B}\right)\geq \xi\Delta^2$ and $\left(\left(1+\left[\frac{\mu^2}{2\tilde{e}\tilde{B}}\right]\right)2\tilde{e}\tilde{B}-\mu^2\right)\geq \xi\Delta^2$ where $\xi$ is a number of order one but greater than one. We plot the expressions for $v_{\perp}$ and $v_{\parallel}$ for all six neutral Goldstone modes as a function of the magnetic field in Fig. \ref{fig:speed}. The empty regions in the curves correspond to a regime where $\left(\mu^2-\left[\frac{\mu^2}{2\tilde{e}\tilde{B}}\right]2\tilde{e}\tilde{B}\right)\leq \xi\Delta^2$ and $\left(\left(1+\left[\frac{\mu^2}{2\tilde{e}\tilde{B}}\right]\right)2\tilde{e}\tilde{B}-\mu^2\right)\leq \xi\Delta^2$ and the above expressions are not valid. In order to determine the regions to be excluded in the plot due the reason mentioned above we have taken $\Delta\sim 0.1\mu$. Although it seems from the regions plotted in Fig. \ref{fig:speed}, that the speed continues to drop to zero as $\left[\frac{\mu^2}{2|\tilde{e}\tilde{B}|}\right]2|\tilde{e}\tilde{B}|\rightarrow\mu^2$ in the regions we have not plotted in the figure, this is not the behavior in reality. It is an artifact of how we computed the integrals for the topmost Landau level($l=l_{max}=\left[\frac{\mu^2}{2|\tilde{e}\tilde{B}|}\right]$) in our calculation. The finite $p_3$ integrals corresponding to the topmost Landau level were performed from $\sqrt{p_3^2+2|\tilde{e}\tilde{B}|l_{max}}-\mu=-\infty$ to $\sqrt{p_3^2+2|\tilde{e}\tilde{B}|l_{max}}-\mu=\infty$ which is a reasonable thing to do when $\left(\mu^2-\left[\frac{\mu^2}{2\tilde{e}\tilde{B}}\right]2\tilde{e}\tilde{B}\right)\geq \xi\Delta^2$ and $\left(\left(1+\left[\frac{\mu^2}{2\tilde{e}\tilde{B}}\right]\right)2\tilde{e}\tilde{B}-\mu^2\right)\geq \xi\Delta^2$  as the the integrands involved go to zero as $\sqrt{p_3^2+2|\tilde{e}\tilde{B}|l_{max}}-\mu>\Delta$. In regions $\left(\mu^2-\left[\frac{\mu^2}{2\tilde{e}\tilde{B}}\right]2\tilde{e}\tilde{B}\right)\leq \xi\Delta^2$ and $\left(\left(1+\left[\frac{\mu^2}{2\tilde{e}\tilde{B}}\right]\right)2\tilde{e}\tilde{B}-\mu^2\right)\leq \xi\Delta^2$ the lower limit of integrals for the topmost landau level should be $p_3=\sqrt{\mu^2-2|\tilde{e}\tilde{B}|l_{max}}$. The implication of this for the Fig. \ref{fig:speed} seems to be that as the magnetic field is increased, the curve for the speed for $l_{max}=l'$ first decrease as $\frac{\mu2}{2\tilde{e}\tilde{B}}\rightarrow l'$ as shown in Fig. \ref{fig:speed} and then increase over a region of $\sim\xi\Delta^2$ about a point where $\frac{\mu2}{2\tilde{e}\tilde{B}}=l'$, which is not captured in Fig. \ref{fig:speed}, to meet the curve for the speed for $l_{max}=l'-1$.
\section{Conclusion}
In the last two sections we analysed the propagator for the neutral Goldstone modes of the color-flavor locked phase in the presence of a magnetic field for both high magnetic fields $\tilde{e}\tilde{B}\sim 2\mu^2$ and more realistic moderate magnetic fields $\Delta^2\leq\tilde{e}\tilde{B}\leq 0.3\mu^2$. We found that the speeds of these modes differ considerably from their values in the absence of a magnetic field found in the weak coupling limit $\sqrt{\frac{1}{3}}$, for the range of the magnetic field we are considering here. There are a couple of improvements that can be made to our calculations here. The first one is for the regime $\tilde{e}\tilde{B}\sim 2\mu^2$. The magnetic field in this regime is so high that only the lowest landau level is filled. This may imply that the magnetic field should already be probing the vacuum which can lead to the condensation of the chiral condensate ($\bar{q}q$). This phenomenon at zero density is known as `magnetic catalysis' \cite{PhysRevD.66.045006,PhysRevLett.73.3499,Gusynin:1994xp,
PhysRevD.52.4747,PhysRevLett.83.1291,Gusynin:1999pq},\cite{PhysRevD.54.4181,PhysRevD.55.6504,Shushpanov:1997sf,PhysRevD.57.3759,Ferrer:2000ed},\cite{
,PhysRevD.62.105017,PhysRevD.66.014004}. This is likely to have an impact on the magnitude of the MCFL gap. If this is the case, then the mesonic excitations will involve quark-antiquark quantum numbers like mesons at zero density along with the diquark quantum numbers considered here. This will in turn add corrections to the anisotropy in the speed of the neutral mesonic excitations that we found here. Also our calculation at $\tilde{e}\tilde{B}\sim 2\mu^2$ involves gaps $\Delta_1=\Delta_3=\Delta$ which is why although we can accurately (barring the corrections coming from the presence of a chiral condensate) predict the anisotropy in the speed of the Goldstone modes at $\tilde{e}\tilde{B}=2\mu^2$, we lose information regarding how this anisotropy changes as a function of $\frac{\tilde{e}\tilde{B}}{\mu^2}$ around the point $\tilde{e}\tilde{B}=2\mu^2$. In order to recover this information the calculation for the propagators of the Goldstone modes needs to be carried out in its full generality using gaps that are unequal or $\Delta_1\neq\Delta_3$. This is not very important at lower magnetic fields $\tilde{e}\tilde{B}<0.3\mu^2$ as the two gaps are close to each other in magnitude in that regime. However, for large magnetic fields $2\tilde{e}\tilde{B}>\mu^2$ this is important as the gaps become vastly different \cite{PhysRevD.76.105030} as the magnetic field is increased from $\tilde{e}\tilde{B}\sim\frac{\mu^2}{2}$ onwards, except for the point $\tilde{e}\tilde{B}\sim 2\mu^2$. A more detailed calculation including the effect of magnetic catalysis and with unequal gaps $\Delta_1$ and $\Delta_3$ will improve the accuracy of this calculation and will also give us a prediction of how the anisotropy in the speed of the Goldstone modes changes as the magnetic field is varied in this regime of very high magnetic field.
\\ As seen in Fig. \ref{fig:speed} the speed of the Goldstone modes not only differs considerably from the standard weak coupling value of $\frac{1}{\sqrt{3}}$ that is found in the absence of a magnetic field but also undergoes de Haas-van Alphen oscillations with increasing magnetic field for fields $\tilde{e}\tilde{B}<\frac{\mu^2}{2}$. These could have observable consequences for the transport properties of the core of a magnetar. The biggest contribution to the neutrino emissivity of the CFL phase comes from the decay of the pseudo Goldstone modes or the scattering of the Goldstone modes into neutrinos. The anisotropy found in the speed of the Goldstone modes here could significantly alter the neutrino emissivity parallel to the magnetic field compared to that perpendicular to it. Also the oscillations observed in the speed could inroduce oscillatory behavior in the neutrino emissivity which could be relevant for a magnetar going through stellar evolution. It was found in \cite{PhysRevD.76.105030} that during its stellar evolution, a magnetar with a CFL core could go through several successive phase transitions which would induce discontinuous changes in the induced magnetic field in the core. This could translate into observable effects in the cooling due to oscillatory nature of the neutrino emission.
\section{Acknowledgements}
The author would like to thank Paulo F. Bedaque and Igor Shovkovy for insightful discussions. This work was supported by the U.S. Department of Energy through grant number DEFG02-93ER-40762.
\section{Appendix}
The Landau level basis has been explained in great detail in \cite{Ferrer:2004yc,Ferrer:2006vw}. Here we outline the basic steps to find the quark propagator for the problem at hand. We only need to find the propagator in the limit $\Delta_1=\Delta_3$. In its current form neither $\Delta^-$ nor its square is diagonal. However our calculations become a lot simpler if the square of the gap matrix is diagonal in the basis we are working in. After we find such a basis for $\Delta^-\big|_{\Delta_1=\Delta_3}$ the condensate in this new basis is given by 
\bea
\Delta' =V \left(\Delta^-\big|_{\Delta_1=\Delta_3}\right) V^t
\eea
where $V$ is the matrix given by
\bea
V =\begin{pmatrix}
-\sqrt{\frac{2}{3}} & 0 & 0 & 0 & \frac{1}{\sqrt{6}} & 0 & 0 & 0 & \frac{1}{\sqrt{6}}\\
0 & 1 & 0 & 0 & 0 & 0 & 0 & 0 & 0\\
0 & 0 & 1 & 0 & 0 & 0 & 0 & 0 & 0\\
0 & 0 & 0 & 1 & 0 & 0 & 0 & 0 & 0\\
0 & 0 & 0 & 0 & \frac{1}{\sqrt{2}} & 0 & 0 & 0 & -\frac{1}{\sqrt{2}}\\
0 & 0 & 0 & 0 & 0 & 1 & 0 & 0 & 0\\
0 & 0 & 0 & 0 & 0 & 0 & 1 & 0 & 0\\
0 & 0 & 0 & 0 & 0 & 0 & 0 & 1 & 0\\
\frac{1}{\sqrt{3}} & 0 & 0 & 0 & \frac{1}{\sqrt{3}} & 0 & 0 & 0 & \frac{1}{\sqrt{3}}\\
\end{pmatrix}
\eea
and $\Delta'$ is,
\bea
\Delta'=\begin{pmatrix}
 -\Delta  & 0 & 0 & 0 & 0 & 0 & 0 & 0 & 0 \\
 0 & 0 & 0 & -\Delta  & 0 & 0 & 0 & 0 & 0 \\
 0 & 0 & 0 & 0 & 0 & 0 & -\Delta  & 0 & 0 \\
 0 & -\Delta  & 0 & 0 & 0 & 0 & 0 & 0 & 0 \\
 0 & 0 & 0 & 0 & -\Delta  & 0 & 0 & 0 & 0 \\
 0 & 0 & 0 & 0 & 0 & 0 & 0 & -\Delta  & 0 \\
 0 & 0 & -\Delta  & 0 & 0 & 0 & 0 & 0 & 0 \\
 0 & 0 & 0 & 0 & 0 & -\Delta  & 0 & 0 & 0 \\
 0 & 0 & 0 & 0 & 0 & 0 & 0 & 0 & 2 \Delta  \\
\end{pmatrix}
\eea with $\Delta_1=\Delta_3=\Delta$.
We need to evaluate the quark propagator in order to proceed. The quark propagator and its inverse are two by two matrices in the Gorkove space,
\bea
S_{\text{quark}}=\begin{pmatrix}
S_{\text{quark}}^{11} & S_{\text{quark}}^{12} \\
S_{\text{quark}}^{21} & S_{\text{quark}}^{22}.
\end{pmatrix}
\eea
$S_{\text{quark}}^{11}$, $S_{\text{quark}}^{12}$, $S_{\text{quark}}^{21}$, $S_{\text{quark}}^{22}$ are all $9\times 9$ matrix in the color flavor space.
The inverse quark propagator $\left(S_{quark}\right)^{-1}$ is diagonal in the Landau level basis and can be expressed as,
\bea
\left(\left(S_{\text{quark}}\right)^{-1}\right)^{11}=((p^0+\mu)\gamma^0 -\sqrt{2\tilde{e}\tilde{B}l}\gamma^2 -p^3\gamma^3)\Omega_{+}+((p^0+\mu)\gamma^0 +\sqrt{2\tilde{e}\tilde{B}l}\gamma^2 -p^3\gamma^3)\Omega_{-}+((p^0+\mu)\gamma^0 -p^3\gamma^3)\Omega_{0}
\eea,
\bea
\left(\left(S_{\text{quark}}\right)^{-1}\right)^{22}=((p^0-\mu)\gamma^0 -\sqrt{2\tilde{e}\tilde{B}l}\gamma^2 -p^3\gamma^3)\Omega_{-}+((p^0-\mu)\gamma^0 +\sqrt{2\tilde{e}\tilde{B}l}\gamma^2 -p^3\gamma^3)\Omega_{+}+((p^0-\mu)\gamma^0 -p^3\gamma^3)\Omega_{0}
\eea,
\bea
\left(\left(S_{\text{quark}}\right)^{-1}\right)^{12}=i\gamma^5\Delta'\equiv \Delta_c'
\eea
and 
\bea
\left(\left(S_{\text{quark}}\right)^{-1}\right)^{21}=\gamma^0(-i\gamma^5\Delta')\gamma^0\equiv\tilde{\Delta}_c'.
\eea
We have defined two matrices $\Delta_c'$ and $\tilde{\Delta}_c'$ as we need them for convenience later in the appendix.
We need to invert $S_{\text{quark}}^{-1}$ to get the quark propagator $S_{\text{quark}}$ in the landau level basis. The inverse of the elements of $S_{\text{quark}}$ are given by
\begin{align}
\begin{split}
(S_{\text{quark}}^{11})^{-1}&= ((p^0+\mu)\gamma^0 -\sqrt{2\tilde{e}\tilde{B}l}\gamma^2 -p^3\gamma^3)\Omega_{+}+((p^0+\mu)\gamma^0 +\sqrt{2\tilde{e}\tilde{B}l}\gamma^2 -p^3\gamma^3)\Omega_{-}+((p^0+\mu)\gamma^0 -p^3\gamma^3)\Omega_{0}\\
&\qquad\quad
-\left(i\gamma^5\Delta'\right)\left[((p^0-\mu)\gamma^0 -\sqrt{2\tilde{e}\tilde{B}l}\gamma^2 -p^3\gamma^3)\Omega_{-}+((p^0-\mu)\gamma^0 +\sqrt{2\tilde{e}\tilde{B}l}\gamma^2 -p^3\gamma^3)\Omega_{+}\right.\\
&\qquad\quad \left.+((p^0-\mu)\gamma^0 -p^3\gamma^3)\Omega_{0}\right]^{-1}\left(\gamma^0(-i\gamma^5)\left(\Delta'\right)^{\dagger}\gamma^0\right)
\end{split}\label{S11}\end{align} ,

\begin{align}
\begin{split}
(S_{\text{quark}}^{22})^{-1}&= ((p^0-\mu)\gamma^0 -\sqrt{2\tilde{e}\tilde{B}l}\gamma^2 -p^3\gamma^3)\Omega_{-}+((p^0-\mu)\gamma^0 +\sqrt{2\tilde{e}\tilde{B}l}\gamma^2 -p^3\gamma^3)\Omega_{+}+((p^0-\mu)\gamma^0 -p^3\gamma^3)\Omega_{0}\\
&\qquad\quad
-\left(\gamma^0(-i\gamma^5)\Delta'^{\dagger}\gamma^0\right)\left[((p^0+\mu)\gamma^0 -\sqrt{2\tilde{e}\tilde{B}l}\gamma^2 -p^3\gamma^3)\Omega_{+}+((p^0+\mu)\gamma^0 +\sqrt{2\tilde{e}\tilde{B}l}\gamma^2 -p^3\gamma^3)\Omega_{-}\right.\\
&\qquad\quad \left.+((p^0+\mu)\gamma^0 -p^3\gamma^3)\Omega_{0}\right]^{-1}(i\gamma^5\Delta')
\end{split}\label{S11}\end{align} ,
\begin{align}
\begin{split}
(S_{\text{quark}}^{12})^{-1}&= -\left[\left[((p^0+\mu)\gamma^0 -\sqrt{2\tilde{e}\tilde{B}l}\gamma^2 -p^3\gamma^3)\Omega_{+}+((p^0+\mu)\gamma^0 +\sqrt{2\tilde{e}\tilde{B}l}\gamma^2 -p^3\gamma^3)\Omega_{-}+((p^0+\mu)\gamma^0 -p^3\gamma^3)\Omega_{0}\right]^{-1}\right.\\
&\qquad\quad \left. \left(i\gamma^5\Delta'\right)S^{22}\right]^{-1}
\end{split}\label{S11}\end{align}
and 
\begin{align}
\begin{split}
(S_{\text{quark}}^{21})^{-1}&= -\left[\left[((p^0-\mu)\gamma^0 -\sqrt{2\tilde{e}\tilde{B}l}\gamma^2 -p^3\gamma^3)\Omega_{-}+((p^0-\mu)\gamma^0 +\sqrt{2\tilde{e}\tilde{B}l}\gamma^2 -p^3\gamma^3)\Omega_{+}+((p^0-\mu)\gamma^0 -p^3\gamma^3)\Omega_{0}\right]^{-1}\right.\\
&\qquad\quad \left. \left(\gamma^0(-i\gamma^5)\Delta'^{\dagger}\gamma^0\right)S^{11}\right]^{-1}.
\end{split}\label{S11}\end{align}
To write out the different components of $S_{\text{quark}}$ explicitly we need to define the following projectors in the massless quarks limit,
  \bea
 \Lambda^{\pm}_{(+)}=\frac{1}{2}\left(1\pm\frac{\gamma^0(\gamma^3p^3 +\gamma^2 \sqrt{2\tilde{e}\tilde{B}l})}{\sqrt{p_3^2+2\tilde{e}\tilde{B}l}}\right)
 \eea
 and
 \bea
 \Lambda^{\pm}_{(-)}=\frac{1}{2}\left(1\pm\frac{\gamma^0(\gamma^3p^3 -\gamma^2 \sqrt{2\tilde{e}\tilde{B}l})}{\sqrt{p_3^2+2\tilde{e}\tilde{B}l}}\right).
 \eea
 From this point onwards we refer to $S_{\text{quark}}$ as $S$. We write the diagonal terms of the propagator in the Gorkov spacep in terms of these projectors as
 \bea
 (S^{11})_{11-11}=S^{11}_{12-12}=S^{11}_{21-21}=S^{11}_{22-22}=\frac{(p_0-|\overline{p}|-\mu)\gamma^0\Lambda^{+}}{p_0^2-(|\overline{p}|+\mu)^2-\Delta^2}+\frac{(p_0+|\overline{p}|-\mu)\gamma^0\Lambda^{-}}{p_0^2-(|\overline{p}|-\mu)^2-\Delta^2},
 \eea
 \bea
 S^{11}_{13-13}=S^{11}_{23-23}=\frac{(p_0-\sqrt{p_3^2+2\tilde{e}\tilde{B}l}-\mu)\gamma^0\Lambda^{+}_{-}}{p_0^2-(\sqrt{p_3^2+2\tilde{e}\tilde{B}l}+\mu)^2-\Delta^2}+\frac{(p_0+\sqrt{p_3^2+
 2\tilde{e}\tilde{B}l}-\mu)\gamma^0\Lambda^{-}_{-}}{p_0^2-(\sqrt{p_3^2+2\tilde{e}\tilde{B}l}-\mu)^2-\Delta^2},
 \eea
 \bea
 S^{11}_{31-31}=S^{11}_{32-32}=\frac{(p_0-\sqrt{p_3^2+2\tilde{e}\tilde{B}l}-\mu)\gamma^0\Lambda^{+}_{+}}{p_0^2-(\sqrt{p_3^2+2\tilde{e}\tilde{B}l}+\mu)^2-\Delta^2}+\frac{(p_0+\sqrt{p_3^2+
 2\tilde{e}\tilde{B}l}-\mu)\gamma^0\Lambda^{-}_{+}}{p_0^2-(\sqrt{p_3^2+2\tilde{e}\tilde{B}l}-\mu)^2-\Delta^2},
 \eea
 \bea
 S^{11}_{33-33}=\frac{(p_0-|\overline{p}|-\mu)\gamma^0\Lambda^{+}}{p_0^2-(|\overline{p}|+\mu)^2-4\Delta^2}+\frac{(p_0+|\overline{p}|-\mu)\gamma^0\Lambda^{-}}{p_0^2-(|\overline{p}|-\mu)^2-4\Delta^2},
 \eea
 and \bea
 S^{22}_{11-11}=S^{22}_{12-12}=S^{22}_{21-21}=S^{22}_{22-22}=\frac{(p_0-|\overline{p}|+\mu)\gamma^0\Lambda^{+}}{p_0^2-(|\overline{p}|-\mu)^2-\Delta^2}+\frac{(p_0+|\overline{p}|+\mu)\gamma^0\Lambda^{-}}{p_0^2-(|\overline{p}|+\mu)^2-\Delta^2},
 \eea
 \bea
 S^{22}_{13-13}=S^{22}_{23-23}=\frac{(p_0-\sqrt{p_3^2+2\tilde{e}\tilde{B}l}+\mu)\gamma^0\Lambda^{+}_{+}}{p_0^2-(\sqrt{p_3^2+2\tilde{e}\tilde{B}l}-\mu)^2-\Delta^2}+\frac{(p_0+\sqrt{p_3^2
 +2\tilde{e}\tilde{B}l}+\mu)\gamma^0\Lambda^{-}_{+}}{p_0^2-(\sqrt{p_3^2+2\tilde{e}\tilde{B}l}+\mu)^2-\Delta^2},
 \eea
 \bea
 S^{22}_{31-31}=S^{22}_{32-32}=\frac{(p_0-\sqrt{p_3^2+2\tilde{e}\tilde{B}l}+\mu)\gamma^0\Lambda^{+}_{-}}{p_0^2-(\sqrt{p_3^2+2\tilde{e}\tilde{B}l}-\mu)^2-\Delta^2}+\frac{(p_0+\sqrt{p_3^2
 +2\tilde{e}\tilde{B}l}+\mu)\gamma^0\Lambda^{-}_{-}}{p_0^2-(\sqrt{p_3^2+2\tilde{e}\tilde{B}l}+\mu)^2-\Delta^2},
 \eea
 \bea
 S^{22}_{33-33}=\frac{(p_0-|\overline{p}|+\mu)\gamma^0\Lambda^{+}}{p_0^2-(|\overline{p}|-\mu)^2-4\Delta^2}+\frac{(p_0+|\overline{p}|+\mu)\gamma^0\Lambda^{-}}{p_0^2-(|\overline{p}|+\mu)^2-4\Delta^2}
 \eea
 where the upper indices on $S$ are Gorkov indices and lower ones are color-flavor indices.
 To write down the expressions for $S^{12}$ and $S^{21}$ with some clarity we need to define the following
 \bea
(Z^{22})^{-1}\equiv\frac{\gamma^0\Lambda^{+}}{p_0+|\overline{p}|-\mu}+\frac{\gamma^0\Lambda^{-}}{p_0-|\overline{p}|-\mu},
\eea
\bea
(M^{22})^{-1}\equiv\frac{\gamma^0\Lambda^{+}_{-}}{p_0+\sqrt{2\tilde{e}\tilde{B}l+p_3^2}-\mu}+\frac{\gamma^0\Lambda^{-}_{-}}{p_0-\sqrt{2\tilde{e}\tilde{B}l+p_3^2}-\mu},
\eea
\bea
(P^{22})^{-1}\equiv\frac{\gamma^0\Lambda^{+}_{+}}{p_0+\sqrt{2\tilde{e}\tilde{B}l+p_3^2}-\mu}+\frac{\gamma^0\Lambda^{-}_{+}}{p_0-\sqrt{2\tilde{e}\tilde{B}l+p_3^2}-\mu},
\eea
\bea
(Z^{11})^{-1}\equiv\frac{\gamma^0\Lambda^{+}}{p_0+|\overline{p}|+\mu}+\frac{\gamma^0\Lambda^{-}}{p_0-|\overline{p}|+\mu},
\eea
\bea
(M^{11})^{-1}\equiv\frac{\gamma^0\Lambda^{+}_{-}}{p_0+\sqrt{2\tilde{e}\tilde{B}l+p_3^2}+\mu}+\frac{\gamma^0\Lambda^{-}_{-}}{p_0-\sqrt{2\tilde{e}\tilde{B}l+p_3^2}+\mu},
\eea
\bea
(P^{11})^{-1}\equiv\frac{\gamma^0\Lambda^{+}_{+}}{p_0+\sqrt{2\tilde{e}\tilde{B}l+p_3^2}+\mu}+\frac{\gamma^0\Lambda^{-}_{+}}{p_0-\sqrt{2\tilde{e}\tilde{B}l+p_3^2}+\mu}.
\eea
 
 $S^{21}$ and $S^{12}$ can be written in terms of these newly defined quantities as
 \tiny
\begin{align}
 &S^{21}=-((S^{-1})^{22})^{-1}(\gamma^{0}(-i\gamma^5)\Delta'^{\dagger}\gamma^0)S^{11}\\
 &=-\begin{pmatrix}
 -(Z^{22})^{-1}\tilde{\Delta}_c'S^{11}_{11} & 0 & 0 & 0 & 0 & 0 & 0 & 0 & 0 \\
 0 & 0 & 0 & -(Z^{22})^{-1}\tilde{\Delta}_c'S^{11}_{21} & 0 & 0 & 0 & 0 & 0 \\
 0 & 0 & 0 & 0 & 0 & 0 & -(P^{22})^{-1}\tilde{\Delta}_c'S^{11}_{31} & 0 & 0 \\
 0 & -(Z^{22})^{-1}\tilde{\Delta}_c'S^{11}_{12} & 0 & 0 & 0 & 0 & 0 & 0 & 0 \\
 0 & 0 & 0 & 0 & -(Z^{22})^{-1}\tilde{\Delta}_c'S^{11}_{22} & 0 & 0 & 0 & 0 \\
 0 & 0 & 0 & 0 & 0 & 0 & 0 & -(P^{22})^{-1}\tilde{\Delta}_c'S^{11}_{32} & 0 \\
 0 & 0 & -(M^{22})^{-1}\tilde{\Delta}_c'S^{11}_{13} & 0 & 0 & 0 & 0 & 0 & 0 \\
 0 & 0 & 0 & 0 & 0 & -(M^{22})^{-1}\tilde{\Delta}_c'S^{11}_{23} & 0 & 0 & 0 \\
 0 & 0 & 0 & 0 & 0 & 0 & 0 & 0 & 2(Z^{22})^{-1}\tilde{\Delta}_c'S^{11}_{33} \\
 \end{pmatrix}
\end{align}
\normalsize
and
 \tiny
\begin{align}
 &S^{12}=-((S^{-1})^{22})^{-1}(i\gamma^5\Delta')S^{11}\\
 &=-\begin{pmatrix}
 -(Z^{11})^{-1}\Delta_c' S^{22}_{11} & 0 & 0 & 0 & 0 & 0 & 0 & 0 & 0 \\
 0 & 0 & 0 & -(Z^{11})^{-1}\Delta_c' S^{22}_{21} & 0 & 0 & 0 & 0 & 0 \\
 0 & 0 & 0 & 0 & 0 & 0 & -(M^{11})^{-1}\Delta_c' S^{22}_{31} & 0 & 0 \\
 0 & -(Z^{11})^{-1}\Delta_c' S^{22}_{12} & 0 & 0 & 0 & 0 & 0 & 0 & 0 \\
 0 & 0 & 0 & 0 & -(Z^{11})^{-1}\Delta_c' S^{22}_{22} & 0 & 0 & 0 & 0 \\
 0 & 0 & 0 & 0 & 0 & 0 & 0 & -(M^{11})^{-1}\Delta_c' S^{22}_{32} & 0 \\
 0 & 0 & -(P^{11})^{-1}\Delta_c' S^{22}_{13} & 0 & 0 & 0 & 0 & 0 & 0 \\
 0 & 0 & 0 & 0 & 0 & -(P^{11})^{-1}\Delta_c' S^{22}_{23} & 0 & 0 & 0 \\
 0 & 0 & 0 & 0 & 0 & 0 & 0 & 0 & 2(Z^{11})^{-1}\Delta_c' S^{22}_{33} \\
 \end{pmatrix}
\end{align}
\normalsize
where $S^{ab}_{ij}$ stands for $S^{ab}_{ij-ij}$.

\bibliographystyle{unsrt}
\bibliography{cfl_mag}
\end{document}